\documentclass[slac_one]{revtex4}
\usepackage{graphicx}
\usepackage{fancyhdr}
\newcommand{\MUnits}{$GeV/c^2$}
\newcommand{\Bot}{$b$}
\newcommand{\ttbar}{$t\bar{t}$}

\newcommand{\ifb}{$fb^{-1}$}

\begin{document}

\title{{\small{Hadron Collider Physics Symposium (HCP2008),
Galena, Illinois, USA}}\\ 
\vspace{12pt}
Top Quark Mass and Cross Section Results from the Tevatron} 

%

\author{F. Garberson, on behalf of the CDF and D0 Collaborations}
\affiliation{UCSB, Santa Barbara, CA 93106, USA}

\begin{abstract}

This note details recent measurements of the top mass and cross section that have been performed at the Tevatron. Most of these measurements have been performed with between 1.0 and 2.0 \ifb\ of data. Basic top physics concepts are discussed, cross section and mass analysis methods and results are presented, and future analysis prospects are discussed with a special attention given to Electroweak determination of the limits on the Standard Model Higgs mass.

\end{abstract}

\maketitle

\thispagestyle{fancy}


\section{\label{sec:Intro}INTRODUCTION}

Since the discovery of the top quark in 1995 by the CDF and D0 Collaborations \cite{bib:CDF_disc} \cite{bib:DO_disc},
thousands of top quark events have been identified at both experiments. This
large sample has allowed for very precise measurements of the mass and cross
section to be performed and cross checked, and tests of expected top properties
such as its charge, spin, lifetime, production, and decay properties are
beginning to become feasible. So far no clear deviation from Standard Model
expectations has been identified. 

\begin{figure*}
\includegraphics[height=2.5in]{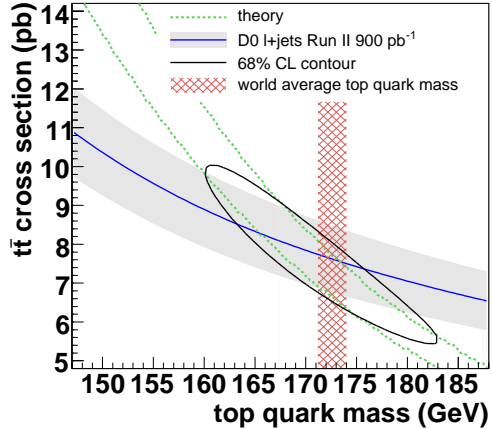}
\caption{Theoretical expectation and uncertainties for the top quark cross section depending on its mass. The D0 Collaboration has used its lepton plus jets cross section measurement and these theoretical predictions to indirectly measure a top quark mass of $170 \pm 7\ GeV/c^2$. }
\label{DO_cross_mass}
\end{figure*}

As precision improves on the cross section measurements it is possible
that an excess will be discovered in some channel that is suggestive of new
physics. And as the precision continues to improve in our mass measurements,
the Electroweak constraints on the Standard Model Higgs mass will continue to
tighten, see Section \ref{sec:Higgs}. Further, a high resolution top mass determination will be useful for calibrating high energy jets at the LHC. But the mass and cross section
measurements should not be considered to be entirely independent of one
another. On the contrary, mass measurements often use the background
distributions identified by cross section measurements as inputs to their
analysis. Perhaps more importantly, kinematic constraints fix the expected top
mass given a cross section. Specifically, the more massive a top quark is, the
more difficult it is to produce at the Tevatron, so the cross section is
expected to fall with increasing mass. The D0 Collaboration has exploited this
fact to use a cross section measurement to indirectly calculate a top mass of $170 \pm 7\ GeV/c^2$\ \cite{bib:DO_lpj_cross},
see Figure \ref{DO_cross_mass}. Details on the cross section and mass measurements are provided later. It is reassuring that this indirect Standard Model
prediction of the relationship between the cross section and the mass is in
agreement with the world average top mass of 172.6 \MUnits.

\section{\label{sec:TopProp}TOP QUARK PROPERTIES}

The top quark is the heaviest known fundamental particle, with a measured mass
of around 172 \MUnits. As a result, due to its unusually large mass, it is likely to play a
key role in any new physics that is discovered at the Electroweak Symmetry
breaking scale. For example, the measured mass of the top quark can be used
along with the measured mass of the W boson to constrain the mass of the
Standard Model Higgs, see Section \ref{sec:Higgs}. Further, since the top width
goes as the cube of the mass to lowest order, the top quark decays extremely
rapidly before it has a chance to hadronize. Thus, it produces a very clean
signature that makes it possible to study its properties
with high precision.

As far as we know, the Standard Model is correct in the prediction that almost
all top quarks decay to \Bot W. The W boson may then decay hadronically or
leptonically. Thus, in \ttbar\ production, three common categories of decays may
occur. If both W's decay to leptons and neutrinos, this is known as the
dilepton channel. Most such analyses require the leptons to be electrons or
muons due to the difficulty of distinguishing taus from jets. As a result this
channel has a small branching fraction of around 5\%. Alternately, one W may
decay leptonically and the other hadronically (to jets), with a larger
branching fraction of 29\%. Finally, both Ws may decay hadronically with a
large branching fraction of about 46\%, however this channel is very
challenging to work in due to the enormous QCD background. Analyses in all of
these channels will be outlined below.

\section{\label{sec:b_tag}\Bot-JET IDENTIFICATION}

In order to achieve a reasonable purity of \ttbar\ events, it is necessary to have a reliable scheme for identifying (tagging) the \Bot-quarks into which they decay. Perhaps the most commonly used approach to \Bot-tagging exploits the characteristic lifetime of the \Bot-hadrons. Since decays of \Bot-hadrons are suppressed according to the CKM matrix, they tend to live longer than charm hadrons, but not so long as strange hadrons. In fact, \Bot-hadrons live roughly long enough that while they will decay before passing outside of the tracking system in most detectors, a displaced secondary decay vertex can be reliably identified using precision silicon tracking. At CDF, the dominant \Bot-tagging algorithm directly exploits this fact by combining displaced tracks into well identified secondary vertices. If these vertices are sufficiently well resolved and displaced from the primary vertex then the jet is considered to be tagged. The D0 Collaboration, on the other hand, has come to prefer a more sophisticated approach. Specifically, they use a neural network that incorporates information about the secondary vertex (decay length, invariant mass, fit quality) as well as information about the track qualities. 

Another approach to \Bot-tagging exploits the significant fraction of \Bot-quarks which decay semileptonically. If a soft muon or electron is found which is associated with a jet, the jet is tagged. This tagging algorithm has a significantly lower efficiency than either of the above taggers, however since it uses complementary information it serves as a cross check to results found with more conventional taggers.

\begin{figure*}
\centerline{
  \mbox{\includegraphics[height=3.0in]{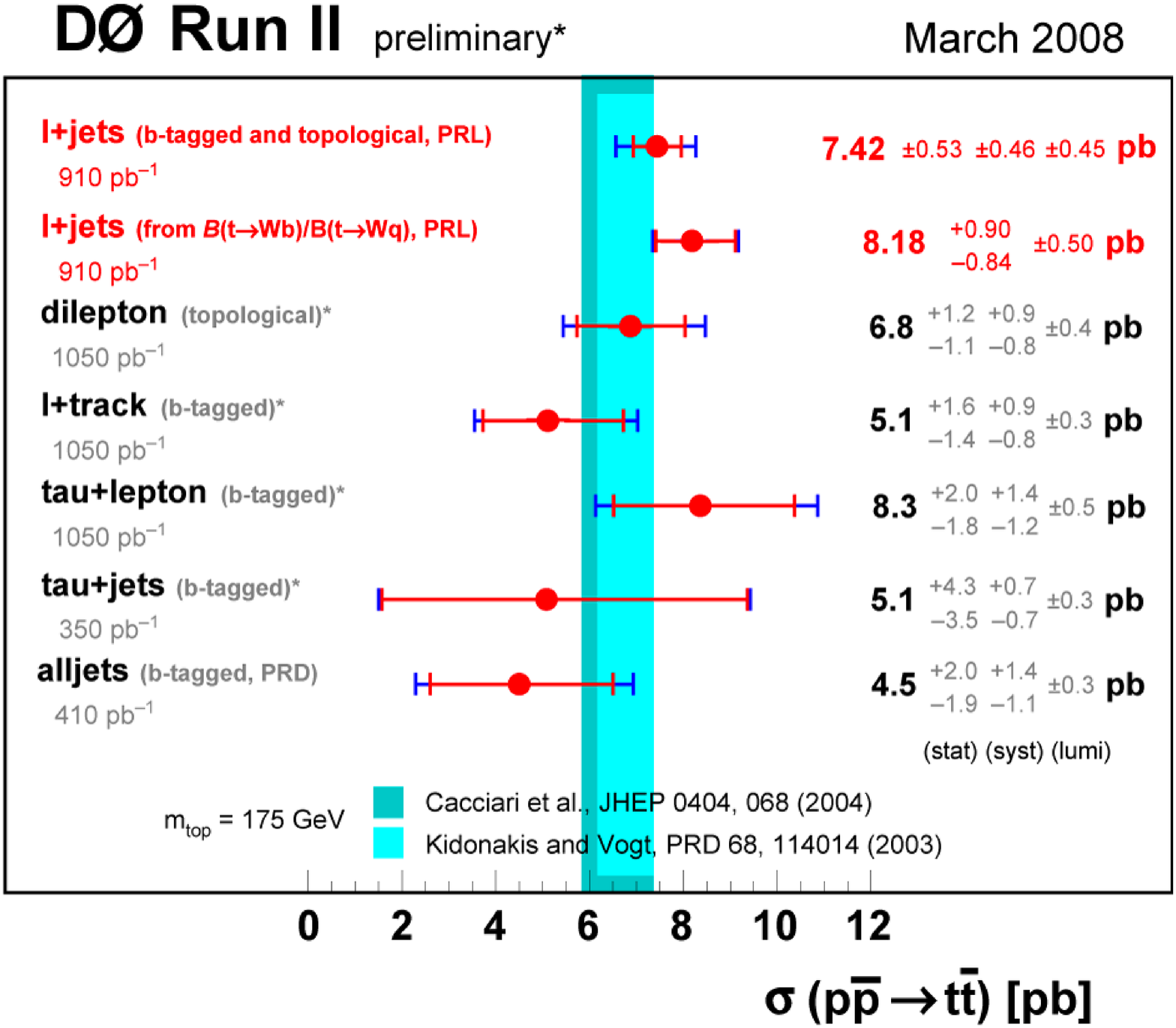}}
  \mbox{\includegraphics[height=4.0in]{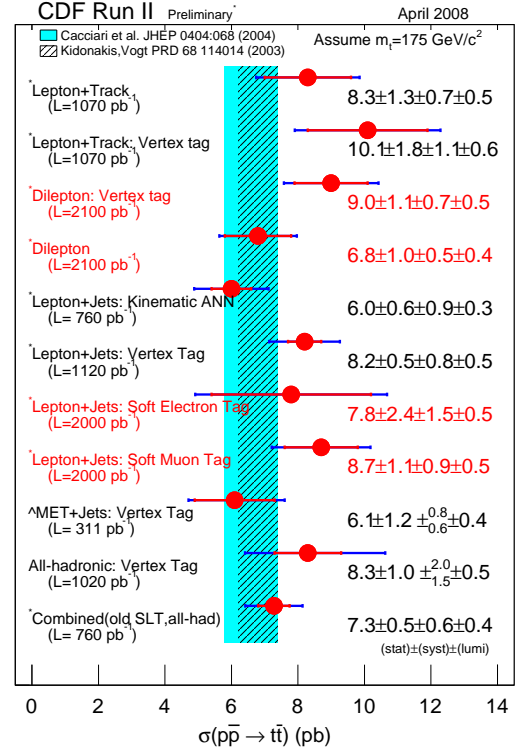}}
}
\caption{Left: Cross section measurements from the D0 Collaboration compared with predictions from theory. Right: Cross section measurements from the CDF Collaboration compared with predictions from theory. }
\label{cross_sect_res}
\end{figure*}

\section{\label{sec:cross}CROSS SECTION MEASUREMENTS}

For a top mass of $175\ GeV/c^2$, the \ttbar\ cross section has been calculated to be about 6.7 pb under the Standard Model at the Tevatron. This prediction along with the many of the recent Tevatron measurements are shown in Figure \ref{cross_sect_res}. Some of the measurements are described below along with references.

\subsection{\label{sec:dil_cross}Cross Section Measurements in the Dilepton Channel}

Events in the dilepton channel are generally distinguished by the
identification of two jets, two isolated leptons, and missing energy from the
two neutrinos. Due to the low branching ratio of the dilepton channel,
statistics have historically been the major limitation of these measurements.
Thus, event selection cuts are often made very loose. At one extreme, no
\Bot-tags are required and one of the two leptons has its identification
requirements loosened so that only a track must be identified. Such an analysis
is known as a "lepton plus track analysis". Even with such loose cuts, a signal
fraction of almost 60\% is seen, but a large background component is
introduced from W+jets where a jet fakes a track, while other sizable backgrounds come from Drell Yan and Diboson production. This analysis has been performed by CDF \cite{bib:cdf_lptrk_cross}.

Other cross section measurements have been performed in the dilepton channel with tighter cuts. The CDF Collaboration has also performed measurements where both leptons must be identified, with \cite{bib:cdf_dil_pretag_cross} and without \cite{bib:cdf_dil_tag_cross} \Bot-tagging. Under these stricter cuts most of the diboson and fakes backgrounds have been eliminated, leaving a signal fraction of greater than 85\%. The D0 Collaboration has measured the dilepton cross section in the lepton plus tracks channel with a \Bot-tagging requirement \cite{bib:DO_lptrk_tag_cross}, and in the dilepton channel without \Bot-tagging \cite{bib:DO_dil_cross}, and combined its results.

\subsection{Cross Section Measurements in the Lepton plus Jets Channel}

The lepton plus jets channel has the advantages both of small backgrounds and of a sizable branching fraction. In identifying such events one commonly requires four jets, an isolated lepton, missing energy from the neutrino, and that at least one of the jets be tagged. When the tagging requirement is imposed, the dominant background comes from W+jets events with a smaller background from QCD production where a jet fakes a lepton and the energy is mismeasured, but the total background fraction will generally be under 30\%. Both CDF \cite{bib:cdf_lpj_tag_cross} and D0 \cite{bib:DO_lpj_cross} have performed analyses using this full selection including \Bot-tagging.

As a cross check to the results determined from \Bot-tagging, both the CDF and
D0 Collaborations have performed measurements without \Bot-tagging
\cite{bib:cdf_lpj_kin_cross} \cite{bib:DO_lpj_cross}. Without tagging, the
backgrounds grow to a sizable fraction of the total events (around 60 or
70\%), so both Collaborations exploit the unique kinematic characteristics of
\ttbar\ production. That is, since top quarks are so massive, they tend to be
produced with minimal boost, and thus decay more centrally than the dominant
backgrounds. CDF trains a neural network and D0 constructs a likelihood
discriminant to separate signal from background. In each case, the
kinematically driven measurements come out consistent with the \Bot-tagging
measurements. D0's combined \Bot-tagging and kinematic results represent the
world's most precise top cross section measurement. Finally, as further cross
checks to the standard measurements, CDF has used alternate soft lepton taggers
to measure the top mass. As explained in Section \ref{sec:b_tag}, these taggers
lack the efficiency of secondary vertex driven taggers, but they exploit
complimentary information to identify the \Bot-quarks. Measurements have been
performed using both electrons \cite{bib:cdf_lpj_soft_ele_cross} and muons
\cite{bib:cdf_lpj_soft_muon_cross} for the tagging, with consistent results.

\subsection{Other Cross Section Measurements}

CDF has performed a cross section measurement in the all hadronic channel \cite{bib:cdf_allhad_cross}. As mentioned previously, this channel is very challenging to work in due to the enormous QCD backgrounds. To select events in this channel, one must search for events with six or more well identified jets with minimal missing energy that pass \Bot-tagging requirements. Even then, the QCD backgrounds are still overwhelming unless one also performs a kinematic based separation between signal and background. Again, CDF trains a neural network to perform this separation, relying on event shape and jet kinematic information. After performing a tight cut on the neural network output, the signal fraction has increased to a more manageable 33\% or so, at which point cross section and mass measurements are performed.

D0 has performed a cross section measurement in dilepton events where one of the leptons is a tau and the tau decays hadronically \cite{bib:DO_tau}. An excess in this channel could potentially be indicative of charged Higgs production. Three categories of tau decays are considered, taus which decay to one track with no EM clusters, taus which decay to one track with an associated EM cluster, and taus which decay to multiple tracks. Neural networks are trained to separate each type of taus from jets based upon shower shape, calorimeter cluster energies, and track qualities. A fourth neural network is additionally trained to separate the second type of tau decay from electrons. The dominant backgrounds of this sample include QCD Multijet events and W+jet events, but a fair amount of signal ends up being from \ttbar\ events in the lepton+jets channel or in the dilepton channel where a tau is misidentified.

\begin{figure*}
\centerline{
  \mbox{\includegraphics[height=4.0in]{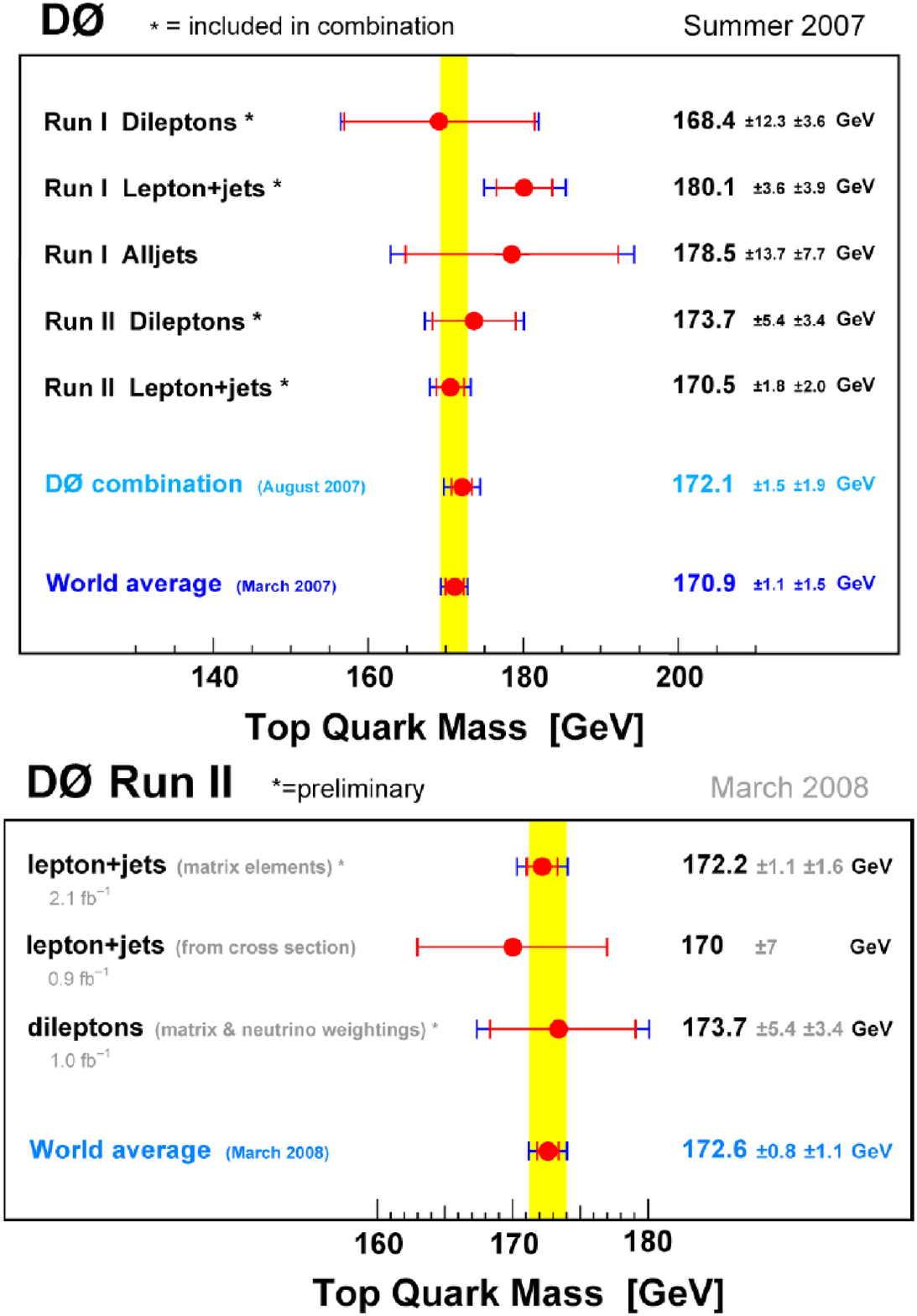}} 
  \mbox{\includegraphics[height=4.0in]{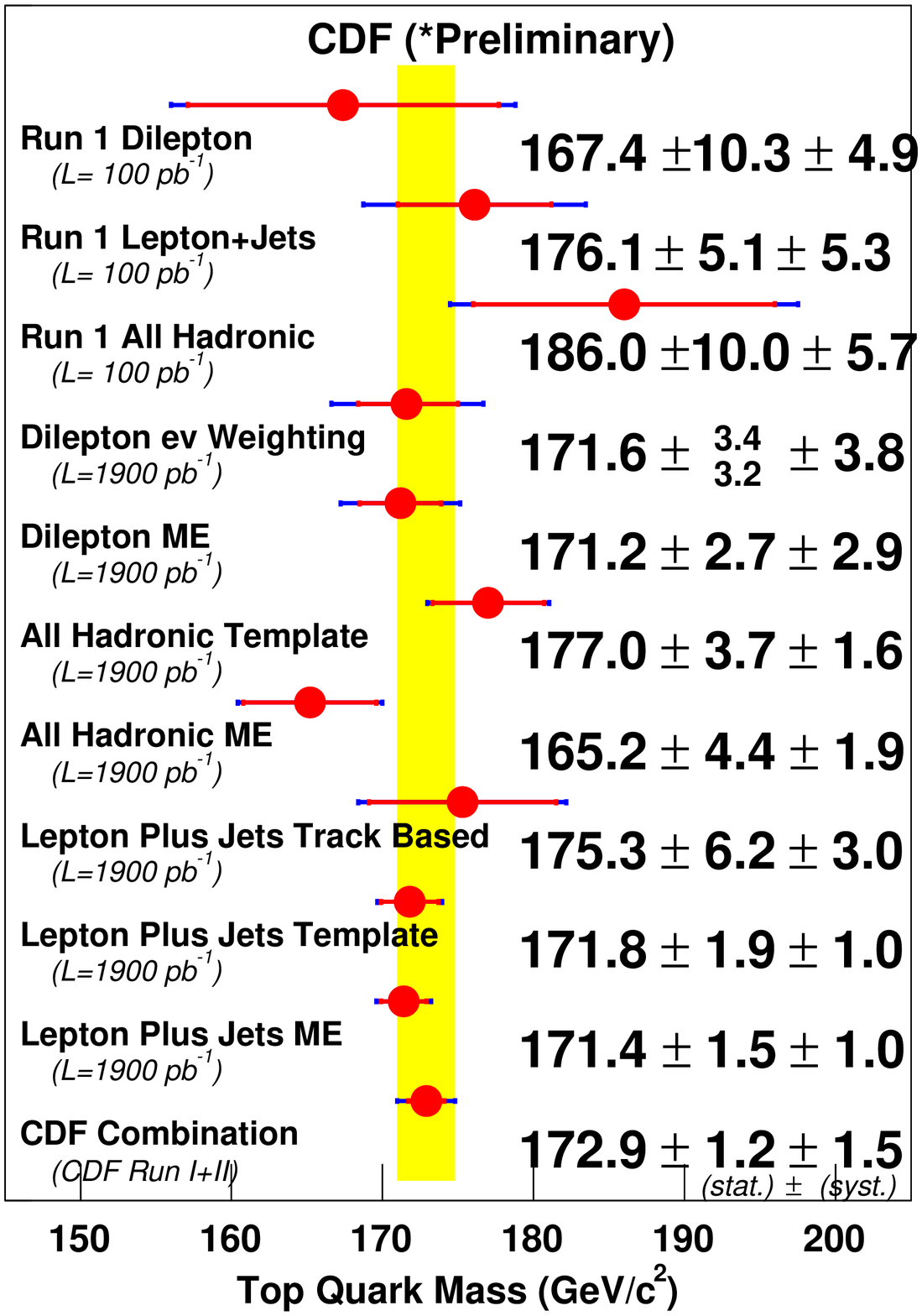}} 
}
\caption{Top left: top mass measurements from the D0 Collaboration and combination as of early 2007. Bottom left: top mass results from the D0 Collaboration since the combination. Right: top mass measurements from the CDF Collaboration.}
\label{mass_res}
\end{figure*}

\section{\label{sec:mass_measurements}MASS MEASUREMENTS}

The results of individual top mass measurements and combinations are shown in
Figure~\ref{mass_res}. Mass analyses must contend with a number of challenges that are less significant for cross
section measurements. One such challenge comes from
combinatorics. The top mass is generally measured by finding the invariant mass
of the four vectors of each of its decay products. It then becomes important to
identify which jets and leptons descend from which quark. The combinatorics
complications are further aggravated when an analysis has to correctly identify
the jets decaying from a W boson, as when an in-situ calibration is applied, see
Subsection \ref{sec:JES}. In practice, sometimes a mass measurement is made for
each combinatoric combination, and the results are combined according to a
likelihood weighting which depends upon the probability of the parton
assignment being correct. Other times only the combination which is calculated to be the most likely to be correct is considered. In addition to combinatorics, another significant complication for mass measurements results from the jet
energy scale correction.

\subsection{\label{sec:JES}Jet Energy Scale}

Most top mass measurements reconstruct the invariant mass of the top
quark from the momenta of its decay products. 
Understanding the momenta of the quarks is not easy because the detector does not
directly measure the momenta of the quarks. Rather it measures the energy of a
spray of particles which descend from the quark according to QCD jet
fragmentation. The uncertainties involved in extrapolating to the correct energy
assignment for each quark based upon the measured jet is known as the jet
energy scale uncertainty, and it is currently the largest uncertainty on the
world average top mass. There are two general approaches one can use for
reducing this uncertainty.

One approach is to "calibrate" the true value of the jet energy scale for each
event (or, more accurately, the proper scale which should be applied to the Monte Carlo to make it agree with data). This calibration exploits the known value of the W boson mass. An
assumption is made that the jet energy corrections are the same for all jets in
the event, and this value is chosen to make the measured W boson mass come out
correct. Obviously this calibration can not be done in dilepton events where
there are no hadronically decaying Ws. Effectively, when one applies this calibration,
the top mass is fitted simultaneously with the jet energy scale. This then converts the uncertainty from a systematic one to a
statistical one. This conversion is not perfect, however. Residual
uncertainties remain from the low order assumption that all jets in the event
should be corrected by the same amount, and other uncertainties arise from
applying the scale derived for light flavor jets from the W decay to the
\Bot-jets in the event. But overall, in the high statistics limit the
systematic uncertainty is expected to drop from more than 3 $GeV/c^2$ to less
than 1 $GeV/c^2$ as a result of this calibration.

Another approach is to circumvent the jet energy scale uncertainty by
not directly using the jet energy information. CDF has performed two mass
measurements using such variables as well as a combination \cite{bib:cdf_TBM}. The variables used are
the mean transverse momentum of leptons from the W boson decays, and
the mean decay length of the \Bot-jets (the \Bot-quark lifetime depends on the
its relativistic boost, which in turn depends on the top mass). Since both
these variables are only weakly correlated to the top mass the statistics must
grow very large for the statistical and some systematic uncertainties to
directly rival conventional mass measurements. But even with only 1.9 \ifb\ at CDF, the
combined measurement has a statistical uncertainty of 6.2 \MUnits\ and a
systematic uncertainty of 3.0 \MUnits. At the LHC these measurements should
become directly competitive, and should have minimal correlation to other
measurements in combination.

\subsection{\label{sec:template} Template Based Mass Measurements}

There are two categories that most top mass measurements can be grouped into.
Many analyses follow the "template method", in which templates are built for
the signal and background samples which are then used in a maximum likelihood
fit on the data. Most recent lepton plus jets and all hadronic analyses perform
these fits in two dimensions, one dimension being the reconstructed top mass,
and the other dimension being the jet energy scale from the in-situ
calibration. 

A number of template based analyses have been performed in the various \ttbar\
decay channels. Some details should be given on the dilepton channel
measurements since they suffer an additional complication due to the ambiguity
in how the two neutrinos are assigned from a single missing energy measurement.
One approach for dealing with this underconstrained system is to use the
\ttbar\ Monte Carlo to find the two dimension probabilities that each neutrino
has a given pseudorapidity. An integration is then performed over kinematically
allowed neutrino direction combinations with this probability applied as a weighting to
perform the mass measurement. CDF \cite{bib:cdf_dil_lpj_comb_temp} and D0 \cite{bib:DO_dil_nuwt} have both performed measurements along
these lines. D0 has also used another approach \cite{bib:DO_dil_matwt} following similar ideas to the
matrix element analyses. In this approach one finds the probability for the
leptons to have the observed momenta as a function of top mass and integrates
over possible parton assignments. The most likely top mass is recorded to build
up signal templates and a likelihood fit is performed to extract a mass
measurement. CDF has also performed template based mass measurements in the
lepton plus jets channel \cite{bib:cdf_dil_lpj_comb_temp} and in the all hadronic channel \cite{bib:cdf_allhad_template}.

\subsection{\label{sec:matrix_element} Matrix Element Based Mass Measurements}

The other major category of analyses follow the "matrix element method". Generally speaking, under this approach one evaluates the likelihoods as a function of top mass for each combinatoric parton permutation, and then combines the likelihood results between permutations and events to determine the final measurement. As an example of such a calculation, the D0 lepton plus jets matrix element analysis \cite{bib:DO_lpj_ME} finds its event by event probability under the signal hypothesis according to the equation below.

\begin{eqnarray*}
\lefteqn{P_{sig}(x, m_{top}, JES) = \frac{1}{\sigma(p\bar{p}\rightarrow t\bar{t}; m_{top}, JES)}} \\
& & \times \sum_{perm} w_i \int_{q_1, q_2, y} \sum_{flavors} dq_1 dq_2 f(q_1)f(q_2) \frac{(2\pi)^4 |M(q\bar{q}\rightarrow t\bar{t} \rightarrow y)|^2}{2q_1q_2s}d\Phi_6W(x,y,JES)
\label{eqn:D0_ME}
\end{eqnarray*}

This equation embodies the following information. If one knew the true four
momenta of all partons, collectively denoted $y$, and the proper matching of
the incoming and outgoing partons perfectly, then the matrix element and cross
section normalization term would be sufficient to determine the likelihood of
observing such an event. In practice one does not know the correct values for the
final states of $y$, one instead measures certain values for the final state,
$x$. The "transfer function", $W$, represents the probability for measuring
$x$, given $y$ and the JES, and must be integrated over all theoretical final
state $y$ values. Integrals over possible initial parton momenta according to
the parton distribution functions complete the conversion from $y$ to $x$.
Finally, one must sum over the possible parton assignments, and weight by the
\Bot-tagging probability, $w$. CDF has also performed matrix element based mass measurements in all three \ttbar\ decay channels \cite{bib:cdf_dil_ME}-\cite{bib:cdf_allhad_ME}.

\subsection{\label{sec:mass_comb}Top Mass Combination and Future Prospects}

A combination of twelve CDF and D0 top mass measurements has recently been
performed \cite{bib:tev_mass_comb}, five from Run-I and seven from Run-II.
Analyses from all decay channels from CDF are combined with analyses from the
lepton plus jets and dilepton channels from D0. Uncertainties on the mass were
divided into twelve presumed orthogonal error categories, six of which detail
uncertainties on the jet energy scale, and others representing uncertainties
due to signal modeling (QCD radiation and parton distribution functions),
background modeling, fitting procedures, and Monte Carlo generation, and the
expected correlation factors between the analyses are accounted for in each
category. With this information as inputs, the analytic BLUE method is used to
evaluate the final mass result and uncertainty of $m_t = 172.6 \pm 1.4\ 
GeV/c^2$, where contributions to the uncertainty are dominated by the jet
energy scale ($\pm0.9$), statistics ($\pm0.8$), signal modeling ($\pm0.5$), and
background modeling ($\pm0.4$).

At the beginning of Run-II at CDF, a goal was set to measure the top
mass to a precision of 3 $GeV/c^2$ by the time 2 \ifb\ of data was accumulated.
This goal has been more than surpassed and the mass is now known with a
certainty of almost 1\%. Since the uncertainty on the mass is no longer
dominated by statistics there may be some concern that further improvements in precision 
can only be minor. Certainly, future prospects will depend largely
on the extent to which the systematic uncertainties can be improved as
illustrated in Figure~\ref{mt_future}. In this figure, CDF makes an
extrapolation of the extent to which the mass precision will improve with
luminosity if systematic improvements are not made (dark blue line), or if
systematic improvements follow the historical trends seen through the first 
\ifb of data by improving at about the rate of statistics (dashed line). This
figure does not take into account improvements seen by combining with the D0
Collaboration.

\begin{figure*}
\includegraphics[height=3.5in]{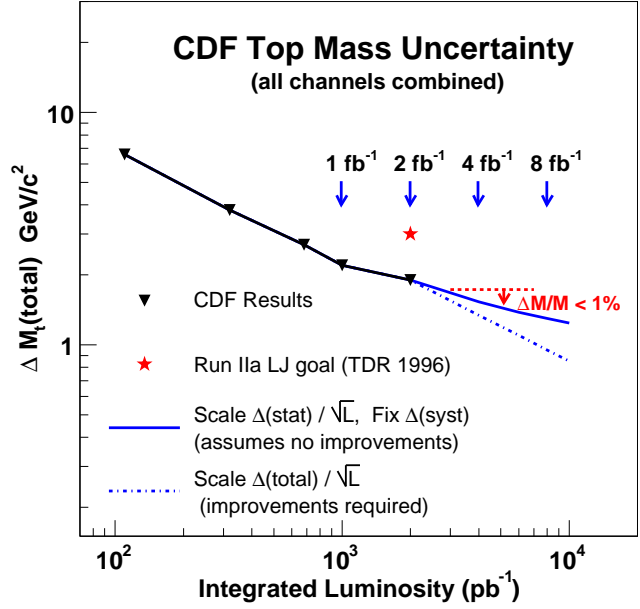}
\caption{Prospects for future improvements in top mass resolution predicted by the CDF Collaboration. Triangles show the historical CDF mass averages, the dark blue line shows the expected rate of mass improvements if no systematic improvements are made, and the dashed blue line shows the expected rate of mass improvements if work is done to improve systematics at the same rate as statistics.}
\label{mt_future}
\end{figure*}

\section{\label{sec:Higgs}HIGGS MASS CONSTRAINTS}

Couplings between the top quark, the W boson, and the Higgs boson (if it
exists) can be used to constrain the Standard Model Higgs mass. Through these
couplings, it can be shown that the Higgs mass depends logarithmically on the
top mass, and quadratically on the W boson mass. Thus, by determining the W
boson mass and the top quark mass, the mass and
uncertainties for the Higgs boson can be indirectly determined. This is
explicitly shown in Figure~\ref{EW_limits}, where the latest results from the most
recent top mass combination and W mass results at the Tevatron are included.
Overall, the Higgs mass is determined to be $87^{+36}_{-27}\ GeV/c^2$, and the
95\% upper bound on the Higgs mass is found to be 160 \MUnits. Note that when
the experimentally determined LEP lower limit on the Higgs mass of 114 \MUnits\ 
is included in the fit, the 95\% upper bound rises to 190 \MUnits.

\begin{figure*}
\centerline{
  \mbox{\includegraphics[height=2.5in]{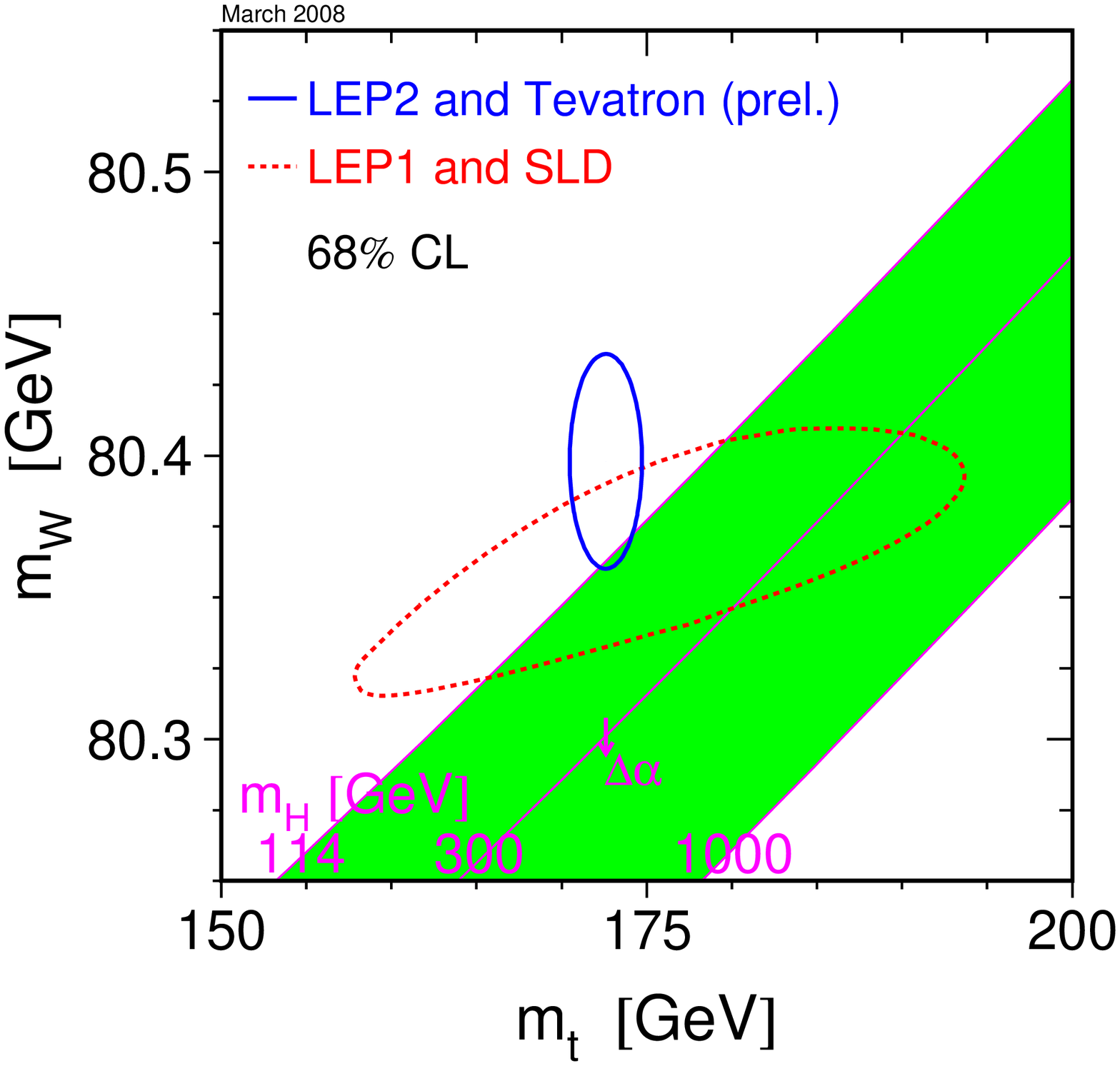}}
  \mbox{\includegraphics[height=2.5in]{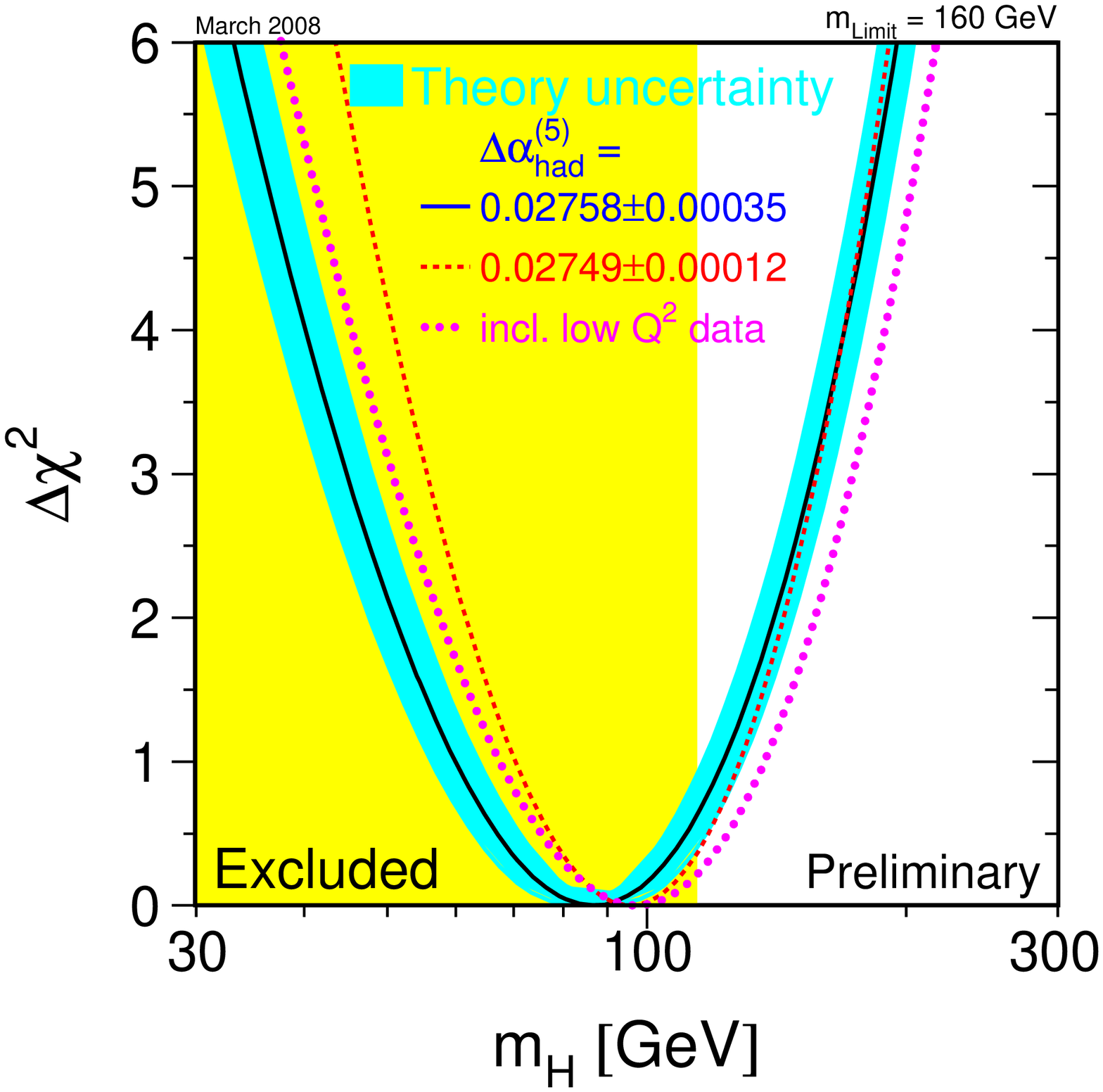}}
}
\caption{Left: Higgs mass constraints depending on most recent measured top mass and W mass results. Right: Results of the global fit to the most likely value of the Standard Model Higgs mass (with some variations based upon different assumed values for the strong coupling constant and which data is included). }
\label{EW_limits}
\end{figure*}

\section{CONCLUSION}

The CDF and D0 Collaborations have measured the \ttbar\ production cross
section in most major production channels and found all to be in agreement with
Standard Model expectations within uncertainties. Thanks to new ideas for
controlling systematic uncertainties they have also measured the top mass to a precision of better than 1\%. Thus far, the mass
measurements and cross section measurements appear to be consistent with one
another, and the mass measurements do not rule out the Standard Model Higgs
boson when applied in conjunction with W boson mass measurements (though a high
mass Higgs is beginning to appear unlikely). When the LHC comes online a new
era of ultra high statistics top quark measurements will become possible
allowing for extremely tight cuts to be made to control backgrounds for cross
section and mass measurements. The top quark mass measurements will also
benefit from the new utility of non-reconstruction based quantities that have
low correlation to traditional measurements \cite{bib:cdf_TBM}. 

The author would like to thank the CDF and D0 Top Quark Physics groups for their
valuable advice, and the HCP organizers for
organizing such an engaging conference and for sheltering us from the rain and
lightning. The author would also like to thank the reader for making it all the way
through to the end of this detailed document.


%

\end{document}